\documentclass[amsmath,amssymb,nofootinbib,prd]{revtex4}

\usepackage{setspace}
\usepackage{graphicx}

\usepackage{amsmath,amssymb,amsfonts,amsthm}
\usepackage{enumerate}

\begin{document}
\title{Quantum panprotopsychism and the combination problem}

\author{Rodolfo Gambini$^1$,  Jorge Pullin\footnote{Corresponding author. Email: pullin@lsu.edu}$^2$}
\affiliation{1. Instituto de F\'{\i}sica, Facultad de Ciencias, Igu\'a 4225, esq. Mataojo,
11400 Montevideo, Uruguay. \\
2. Department of Physics and Astronomy, Louisiana State University,
Baton Rouge, LA 70803-4001, USA.}

\begin{abstract}
We will argue that a phenomenological analysis of consciousness, similar to that of Husserl, shows that the effects of phenomenal qualities shape our perception of the world. It also shows the way the physical and mathematical sciences operate, allowing us to accurately describe the observed regularities in terms of communicable mathematical laws. The latter say nothing about the intrinsic features of things. They only refer to the observed regularities in their behaviors, providing rigorous descriptions of how the universe works, to which any viable ontology must conform. Classical mechanistic determinism limits everything that can occur to what happens in an instant and leaves no room for novelty or any intrinsic aspect that is not epiphenomenal. The situation changes with quantum probabilistic determinism if one takes seriously the ontology that arises from its axioms of objects, systems in certain states, and the events they produce in other objects. As Bertrand Russell pointed out almost a century ago, an ontology of events, with an internal phenomenal aspect, now known as panprotopsychism, is better suited to explaining the phenomenal aspects of consciousness. The central observation of this paper is that many objections to panpsychism and panprotopsychism, which are usually called the combination problem, arise from implicit hypotheses based on classical physics about supervenience. These are inappropriate at the quantum level, where an exponential number of emergent properties and states arise.  The analysis imposes conditions on the possible implementations of quantum cognition mechanisms in the brain.
\end{abstract}
\maketitle

\section{Introduction}

Perceiving, thinking, feeling, believing, or desiring are common mental phenomena shared by all humans that depend on our conscious experiences in some way. We access consciousness directly and through it, everything else that we know. However, it is particularly challenging to reflect on it. Although our current scientific understanding of nature has come to us through consciousness, we encounter significant difficulties when we try to define it or understand its meaning from the natural sciences. Although life without consciousness would be meaningless, we try to minimize its importance and consider it as an emergent phenomenon that will eventually be explained as just another physical process. Incorporating consciousness into our scientific description of natural reality, i.e., solving what Chalmers \cite{chalmers} calls the ``hard problem of mind-body", is not possible without first establishing what type of information the physical sciences provide about the material nature of the world.

To analyze consciousness in scientific terms it is necessary to understand what science tells us about nature, and to do so it is essential to start from a phenomenological approach, that is, from the only direct way to access the world. Everyone has conscious experiences that manifest in various forms in sounds, colors, smells, or feelings of pain or joy. They all have qualitative, personal and non-communicable manifestations that are generally called phenomenal qualities or qualia. These are the qualities to which we have first-person access. Husserl \cite{ideas} developed a discipline that deals with our conscious experiences, which he called phenomenology. For Husserl, following Brentano \cite{brentano}, an essential aspect of consciousness is its intentionality. In general terms, being conscious means being aware of something. He refers to intentionality as this essential characteristic of every conscious act: it is directed towards an object, whether it is a physical object, imaginary, or merely a hallucination. Husserl \cite{ideas} refers to intentionality as a complex of relations that includes us as subjects and also includes the perceptive act, its conceptualization in what he calls noema, and the natural or cultural object that generates it. Intentionality is only the first manifestation of consciousness that reflects on itself. But the task of phenomenology is more general; it consists of reflecting on the phenomena that occur in our consciousness, leaving aside their physical nature or their cultural origin. The possibility of phenomenological reflection ---analyzing consciousness in action--- highlights, as we will see, our conscious contents, the self-consciousness we can have of the same, and frees us from the abstractions with which we often replace them. Explaining the phenomenal qualities that phenomenology deals with in terms of brain processes has proved to be very difficult until now. One of the oldest attempts that continues to receive much attention is panpsychism. Our goal here is to show that a substantial advance to the panpsychist approach is possible from a revision of the philosophical assumptions largely inspired by classical physics and its appropriateness to modern physics. It has been a century since the foundations of quantum mechanics were established, and their inclusion in philosophical analysis has been sporadic. This has been justified based on arguments that rely on the inapplicability of quantum mechanics to certain aspects of reality and the risk that entails using it without rigor. In this work, our aim is to show that an analysis of the problem of consciousness in quantum terms is possible and resolves fundamental objections to certain approaches, such as panpsychism.

In Section II, we will analyze phenomenologically different aspects of our conscious experiences: their intentionality, temporality, self-identity, and our vision of the world and the bodies that occupy and move within it. All of these manifestations of consciousness demonstrate the effectiveness of phenomenal qualities that are both analysable and that shape our perception of the world. We will discuss the limits of the physical-mathematical description, as initially formulated by Galileo and later adapted by psychology and neuroscience to the analysis of the mind-body problem. In Section III, we will present a philosophical view compatible with physics that considers the ingredients of consciousness to be present in most of the natural world. This is a version of Russell's panprotopsychism that results from an ontology based on quantum mechanics that differs significantly from those proposed so far and allows many of the subproblems of the combination problem to be solved. This consists of understanding how the protophenomenal aspects associated with the fundamental entities, say elementary particles, can combine to give rise to our conscious experiences. The traditional formulation of the combination problem is often based on philosophical assumptions that result from a notion of supervenience based on local properties and states. A notion strongly rooted in and inspired by classical physics that is inappropriate for describing the emergence of quantum states and properties that, as we shall see, are more suitable for understanding the phenomenal aspects of consciousness. In Section IV, we will analyze in detail the emergence of states and properties in quantum systems and show that such a notion solves many aspects of the combination problem and makes it essential to include a quantum system that interacts with the neural one. Finally, we will mention several examples showing that quantum mechanics can manifest itself without its effects being destroyed by decoherence in biological systems and the brain. We will conclude in Section V by presenting a summary view of the results of the analysis of the phenomenal dimension of consciousness in terms of quantum panprotopsychism.

\section{Husserl: Phenomenology of the conscious experience}

The task of describing consciousness in action is extremely difficult. We are so immersed in our projects and work, in our affections and fears, that the basic functions of our mind remain implicit and any attempt at analysis is incomplete and falls into contradictions. For his analysis of intentionality, Husserl \cite{husserl} was forced to proceed with extreme precision in a field where every sentence runs the risk of being misinterpreted. Indeed, due to its own nature, consciousness tends to pass unnoticed and get lost behind the object to which it is directed. We live absorbed in things, and when we direct ourselves toward our own conscious acts to follow our intentional behavior and even to identify our phenomenal contents and understand the original perceptual stratum \cite{Waelhens}, we do it in an imprecise or simplistic way.

\subsection{Intentionality}

Intentionality is the most widespread characteristic of consciousness: every conscious act is an attribution of meaning, it is the consciousness of something. That process of directing oneself towards something and attributing it a meaning is what Husserl calls intentionality. It is our most common form of consciousness and accompanies our basic actions of looking, hearing, or touching, and it occurs without the need for perceptions to be accompanied by an object that actually exists. Most of our intentional acts go unnoticed, and the conscious act is not always accompanied by our self-awareness of this. The attribution of meaning in perception, Husserl calls noesis. The attributed meaning has a conceptual character and is called a noema. For the same object, different noemas can be attributed depending on the perspective, the time and the distance. We are conscious of objects, but not neutrally but from structures of meaning that we attribute to them. The different aspects of the intentional process occur as a whole; when I see my dog Ruby entering the garden, there is no process that occurs in stages: an animal, a dog, my dog Ruby. It is the global reality that is given: in this case <my dog Ruby entering the garden>, using Husserl's notation \cite{husserl}. The noema, the statement that accompanies the object, is not an accessory of the thing, but the synthesis of those experiences that are most significant in this moment of my intentional life. Therefore, the assignment of meaning depends essentially on who performs it. Noesis is a meaning assignment performed by a ``me": ``There is no meaning for me without my essential participation" \cite{robberechts}.

As shown, for example, by optical illusions that attribute color to gray regions or volume to flat drawings, the persistent activity of intentionality prioritizes the noema over the phenomenal qualities that appear, as noted by Merleau-Ponty \cite{merleau}, in a horizon of meaning of what is perceived. Perceptions are not combinations of sensations, but they precede them. As he puts it: ``The parts of a thing are not linked together by a simple external association that would be the result of their interdependence... At first, I see certain wholes that I have never seen before: such as houses, the sun, or mountains.'' Conscious activity is primarily animated by the intention of attributing meaning to the world that reaches our senses.

Those meanings do not appear mixed or arbitrarily collected, but as a single context in which those individual meanings are contained. Paraphrasing Husserl: I am sitting in the room in front of the computer and I am aware of the furniture, books, lights, etc. I am also aware of the space behind me and the sounds coming from a nearby construction, the blue sky I see through the window, the current morning moment, the recent breakfast, and the day's tasks that I have to perform. The modes of manifestation of consciousness should be understood not as psychic occurrences similar to physical ones but as meaningful experiential engagements in their own right, that take place in the context of a world of intentional implications and motivations.

\subsection{Time and the temporality of the self}

Something similar happens with time; we do not perceive time as a succession of instants. We do not perceive movement as a succession of positions, our now has a certain thickness. When listening to a melody, we do not hear an instantaneous note; we hear the entire sequence as if the present now retains the past. Something similar happens when we observe the leaves of a tree stirred by the wind and contemplate a dance more than a succession of positions. Husserl refers to retention as the process that preserves what just happened and projection to the anticipation of the future. We understand language thanks to this process. If someone tells us that ``The wind blows from the west" at the moment we hear the ``from", the beginning of the sentence that starts the phrase remains present, and we already have the expectation of a region or cardinal point that completes it. The sentence presents itself as coexisting in the same now and along with it the certainty of being I who listen to it. Our temporality, our inevitable advancement of the now that preserves the past and anticipates the future, is not a succession, but a transition. We do not rest in the present, now we wait for the future, we retain the past and traverse the present. ``Time... is the texture itself of my life."\cite{robberechts}. Thinking about our own temporality and the inevitable advance of the "nows" we become aware of our personal finitude.  But our temporality is not closed on ourselves, the now, insofar as we are with others, it is the common time of who we are in the world. 

We have not referred here to another time than the phenomenological one. This analysis neglects that concepts such as simultaneity are radically modified when the scales of certain physical quantities depart from the ordinary, which could change the usual way of experiencing time under such conditions.

\subsection{Self-consciousness}

Self-consciousness in the sense used here is consciousness of oneself and of our phenomenal contents. The issue of the temporality of the self has been an object of philosophical reflection that also requires a phenomenological approach.  We assume the sameness of consciousness across time because we think in courses of action that will be followed by us. Locke \cite{locke} was the first to notice the temporal character of the self ``For it is by the consciousness it has of its present thoughts and actions that it is self to itself now, and so will be the same self, as long as the same consciousness can extend to actions past or to come...'' This continuity of the self seems to be fundamental for the understanding of our efforts for self-preservation and self-development. The simplest explanation to this is to consider that the unity of the different selves at different times corresponds to the same human person and that the bearer of the first-person perspectives at different times is the same human person including her memories and anticipations. But what Locke is demanding essentially is ``a soul which is at the judgment day all that it ever was'', according to William James \cite{james}. 

Strawson \cite{strawson} and other analytic philosophers consider the existence of a continuous ``me" to be an illusion. For him the remembered past and anticipated future are events in the life of the human being he is. Strawson's point of view makes however very difficult to understand a good fraction of our deepest behaviors and beliefs. As we say in \cite{persons}: How to understand in this hypothesis that we permanently sacrifice for future goals, that we choose a career or devote ourselves to a lifestyle? How to understand the athlete that trains, the tennis player who suffers when she has no energy to climb the ranking or win a tournament without a ``me'' that thinks about a future reward? How to understand feelings that go from the mere pursuit of pleasure to the fear to intense pain that one may suffer in a future moment or to death? Is it enough as an explanation to say one worries about what will happen in the future because it will be experienced by a self that will take care of the human being, also not identical to the present one, that will exist in the future?

Kierkegaard was one of those who best captured the idea that the self has a form of permanence that plays an essential role at the time of understanding how the feeling of responsibility emerges and conditions our behavior: ``For Kierkegaard, ...I don’t take moral responsibility for myself based on a cognitive or reflective apprehension of identity with my remembered past, but rather that experience of identity is itself an experience of responsibility. I am able to become contemporary with temporally distant events precisely insofar as I experience them as something that claims, obligates or ‘calls to’ me as an agent, and because this is a non-reflective, ‘inward’ experience'' \cite{stokes}.

The essential temporality of the self has been recognized on many occasions ``As human satisfaction is different from animal and far surpassing its scope, so is human suffering... only man can be happy or unhappy, thanks to the measuring of his being against terms that transcend the immediate situation. Supremely concerned with what he is, how he lives, what he makes out of himself, and viewing himself from the distance of his wishes, aspirations and approvals, man and man alone is open to despair'' \cite{jonas}. The observation of Jonas is related to the very notion of time as a transition that we have discussed above, which also points to the persistence of the self. 

\subsection{Mathematical and physical sciences and their access to the natural world}

The modern conception of empirical science emerged with Copernican heliocentrism. Nature operates uniformly and universally at all times and in all regions of space. It is governed by mathematically precise laws, which are not always accessible to direct perception but always account for observable data. Kepler, guided by an unshakable faith in the mathematical ordering of events, generated geometric hypotheses about the nature of planetary orbits. He constructs ``on this basis a calculation, i.e., derives from them the movements through rigorous deductive tests." \cite{kepler}.

Husserl attributes to Galileo having made nature more tractable in terms of mathematics by introducing the distinction between primary and secondary qualities. Galileo says ``tastes, odors, colors and so on... reside only in the consciousness... I think that nothing is required in external bodies except shapes, numbers and slow or rapid movements. I think that if ears, noses, and tongues were removed, shapes and numbers and motions would remain, but not odors or tastes or sounds'' \cite{saggiatore}. Galileo is too astute not to notice that both the primary and secondary qualities are mental. The attribution of mental qualities to ``external bodies" involves an abstraction. For Husserl, its origin lies in the perfect transferability of mathematical objects. Implicitly following Plato, he considers that the mathematical world can be built as something that has ``being in itself". However, qualia are intransferable and subjective. Mathematical concepts, for example, the triangle, are shared by everyone who has acquired the basics of a geometrical education. But this mathematization implies a danger: that of identifying the world of external bodies with the conceptual description of their regular behaviors. Husserl \cite{crisis} considers that the abstraction initially produced for the physical world has become extensive to the entire scientific approach, including efforts to understand the mental in terms of the neurophysical. The identification of nature with a mere mathematical description has recently been taken to the extreme by Tegmark \cite{tegmark} who proposed that our universe could satisfy the following hypothesis, called Mathematical Universe Hypothesis (MUH): ``Our external physical reality is a mathematical structure." He distinguishes between ``the outside view or bird perspective of a mathematician studying the mathematical structure and the inside view or frog perspective of an observer living in it." This would be an extreme case where the formal structure, the mathematical laws, precedes and ultimately determines how to compute the inside view, particularly the one analyzed by phenomenology from the outside view.

The remarkable scientific success of this view, based on an abstraction that tends to identify nature with its mathematical description and assumes that such a description is compatible with a mind that perceives, feels, and reasons, led to the ruin of philosophical thought. It became trapped in false alternatives, idealistic, materialistic, or dualistic by what Whitehead \cite{whitehead} calls ``the ascription of misplaced concreteness" that arises from confusing the nature of physical objects with their mathematical description.

\subsection{Regularism}

Physicists continue to operate today with the same confidence that the objective universe ---that is, the accessible in third person--- presents mathematical regularities. The task of scientists is the conceptual representation of these regularities. Husserl \cite{crisis} considers that science entails a danger that has been intensified over time: replacing the concrete world that reveals itself in our experiences with its mathematical idealization. With Husserl and many contemporary physicists and philosophers, we consider that mathematical physics is a description in conceptual terms of the regularities found in our experiences of nature. We call regularism \cite{hospitable} that position, shared by others like Wittgenstein and Hawking. Physics provides information about a natural world that transcends it.

According to Wittgenstein, physical laws have a descriptive and not an explanatory character: ``The whole modern conception of the world is founded on the illusion that the so-called laws of nature are explanations of natural phenomena'' \cite{wittgensteintractat}. The limitation that naturalism legitimately establishes is that descriptions of natural phenomena that imply violations of the laws of physics cannot be admissible. Stephen Hawking \cite{hawking} also agreed with regularism in the sense defined above. He says ``Even if there is only one possible unified theory, it is just a set of rules and equations. What is it that breathes fire into the equations and makes a universe for them to describe? The usual approach of science of constructing a mathematical model cannot answer the question of why there should be a universe for the model to describe.'' This distinguishes with total clarity between the universe and the laws that describe it. Regularism is limited to establishing that the regularities that physics identifies in its description of processes and phenomena are fulfilled without exception, without asserting that nature is exhausted by the physical description. There is, as we will see, a fundamental difference between a regularism based on classical physics and one based on quantum mechanics, since in classical physics there is causal closure, and in quantum mechanics not. Present events are totally determined by the past, while in a quantum world, such as the one we inhabit, determination is only probabilistic.

Note that Hawking, like the vast majority of physicists, adopts some form of realism, a viewpoint that, despite the use of terminology that may lead to misunderstandings, is also shared by Husserl. Usually, it is understood that realism is the thesis that the world around us exists independently of our perception or thinking about it, and idealism is the opposite thesis, of which an extreme view is that of Berkeley. Despite declaring himself a transcendental idealist, in regard to the discussion between realism and idealism, Edmund Husserl does not adhere to any traditional form of idealism. Instead, his position somewhat approaches epistemological realism, and his phenomenological work based on the notion of intentionality reinforces this viewpoint. Husserl justifies his use of the term ``idealism" as follows: ``talking about 'idealism' is not, of course, talking about a metaphysical doctrine but about a theory of knowledge that recognizes the 'ideal' as a condition for the possibility of general objective knowledge, and does not reject it psychologically" \cite{husserllogic}. ``The essence of transcendental idealism for Husserl was the a priori correlation between objectivity and subjectivity" \cite{dictionary}. ``The world is the correlate of consciousness". Although he emphasizes that consciousness is the primary source of all evidence, the transcendental idealism of Husserl, based on the notion of intentionality, allows for direct contact with reality, which we take for granted in our usual behavior.

\subsection{Life-world, art, and freedom}

In his later works, Husserl expanded the range of his interests, and in addition to dealing with our consciousness and its basic behavior, he introduced the concept of life-world, which recognizes that even at its deepest level, consciousness is already operating in a world of meanings and prejudices that are socially, culturally, and historically constituted through an intersubjective process that begins at birth. Our life takes place in time, but within a historical and social context given by what Husserl called the life-world. Each person builds, depending on their specific circumstances and decisions, their own personal life-world within a broader life-world of the society in which we live. Husserl is not clear on this point, he speaks of a plurality of life-worlds, but also emphasizes that in the end there is only one world. One could say that with this concept, Husserl opens a sphere of study hardly explored that involves everything from the implicit foundations of natural sciences to the set of assumptions with which we conduct our daily life. What is immediately given or assumed in our behavior. Without a set of assumptions, there is no possible experience. As Wittgenstein observes, even systematic doubt is not possible \cite{wittgenstein2}. With the term life-world, the context of human knowledge and behavior that cannot be completely analyzed and is permanently active in humans ``even when they are absorbed in the practice of science" is denoted. It is ultimately ``the world of everyday experience" \cite{crisis}. Living is always done in the implicit acceptance of a world, ``to live is always to live in certainty of the world" which although implicit provides the context and meaning of our experiences.

The life-world provides a context that is always present in any experience. Husserl justly opposes this world to the one described by the natural sciences. He says ``What we take as things are pictures, statues, gardens, houses, tables, clothes, tools, etc. They are value objects, use objects, practical objects. They are not objects that can be found in natural science." \cite{ideas}. A person's life-world is built depending on their particular life conditions. It is the world of our interests, purposes, efforts, capabilities, and habits. Of our searches and achievements. The world in which we act and suffer. It is a subjective world and to a large extent inter-subjective that does not include isolated natural objects separated from their cultural and human significance. It is a world loaded with value, sense, and meaning. The life-world is a result of our perceptual intuition. As Husserl says, perception is the ``primary mode of intuition". Every human activity is founded in the ``passively having the world of perceptual intuition. Without such intuition human beings would not have objects of whatever kind they could refer to and be interested in". Summarizing, the life-world is ``the self-evidently existing, evermore intuitively given world" \cite{crisis}. Art is the most obvious objective manifestation of the life-world and as such not only expresses it, but also contributes to its intersubjective construction. Art allows for the expression of what already existed implicitly in the subject and to make it visible and objective. Artistic work always implies a choice in the life-world area only possible thanks to our ability to select, abstract, and conceptualize. Hans Jonas \cite{jonas} notes that freedom is what determines man’s ``specific difference" in the animal kingdom. Freedom is at the heart of any human creation, be it a tool, an image, or a new concept. When we observe a cave painting ``that must have been produced artificially, that have no structural function, and that suggest a likeness to one or another of the living forms encountered outside" we are invaded by the certainty of its human origin. It does not matter how imperfect it is, we think that there was a ``speaking, thinking, inventing, in short 'symbolical'" being. The image abstracts and stylizes essential elements of the object. It chooses to emphasize some and leave out others completely but is created with the intention of being recognizable and identifiable. It is therefore directed to another being that shares with the artist an essential representational ability. The most relevant form of artistic expression is given in the masterpiece in what Nietzsche calls` `the grand style". 

The artist has a purpose at the beginning that prefigures the final result not in its detailed implementation, but in concepts, feelings, and emotions that arise from his life-world and want to be expressed in the work. The religiousness of Greek sculpture in the statue of Hermes of Praxiteles carrying the Infant Dionysius, the world of a peasant woman in northern Europe in Van Gogh's shoes that Heidegger analyzes in ``The Origin of the Work of Art", the horror of the Spanish Civil war in Picasso's Guernica.

Graphic documentation of Picasso painting shows that his creative process includes many times trials that he considers failed and corrections or simply deletion of parts of the work. That is what happened with the preparation of Guernica, of which hundreds of preparatory sketches are preserved and show that parts of the final painted work were erased and repainted. This process shows that the artist creates his work through a careful process of reflection, in which he judges and chooses to keep or eliminate partial representations of his work. It is a process in which the work is constructed through a series of judgments of appropriateness to certain criteria implicitly embedded in his life-world. The choice of allegorical figures and other expressive means to express the horror and barbarity of war exemplifies aspects of every creative process that involves a sequence of choices between various options, evaluations, and actions. Paraphrasing Foucault, one would say that not only is ``freedom the ontological condition of ethics" but also of aesthetics \cite{ethics}.

If quantum indeterminacy could be involved in our mental functioning, the regularism advanced above could allow our freedom to be something more than random \cite{persons}. But for that, it is necessary to identify a form of ethical and aesthetic orientation to which our consciousness has access, compatible with the current vision of our being in the world. We do not propose here to give an answer to this problem. We will only observe that not only are we capable of posing a universe of possible aesthetic or moral options but also valuing them as more or less suitable for the situation, with a certain degree of accuracy, and choosing in many cases to adjust to those valuations. This capacity seems to have an intuitive foundation, more similar to a form of perception than to rational analysis. Christine Korsgaard \cite{korsgaard} argues in this sense that ``The capacity for self-conscious reflection about our own actions confers on us a kind of authority over ourselves, and it is this authority which gives normativity to moral claims" and adds that it is a conception of the right and the good ``under which you value yourself and find your life to be worth living." 

\subsection{Consciousness effectiveness}

Considerations such as the preceding make it necessary to recognize the effectiveness of phenomenal consciousness in presenting analyzable phenomena that, far from being ineffable and intransferable, are common to all people. Like the outer world, consciousness is open to conceptual analysis when appropriate concepts are introduced. This does not result from a process of naturalization or reduction of the phenomenological, but rather highlights an independent domain of the intersubjective reality through what Husserl calls an ``objective science of the spirit" \cite{vienna}. If, on the contrary, the dominant physicalism of these times were correct and nothing could escape becoming governed by physical laws, and if, against all the phenomenological evidence, the ultimate nature of the world were purely physico-mathematical, there would be no space for the exercise of responsible and creative freedom, and the indeterminacy allowed by quantum theories would only yield ``aimless random processes" \cite{kant}. In that case, the crisis denounced by Husserl \cite{crisis} could only lead to the abandonment of finding solid foundations for moral and artistic action and considering freedom and consciousness as illusory.

\section{Panprotopsychism}

\subsection{Events and the third person description of the physical world}

The analysis of consciousness requires distinguishing between third- and first-person perspectives. Science, particularly physics, describes the world from a third-person perspective. As we have seen, the mathematical description provided by the physical sciences establishes the observed regularities that are compatible with other observers and transmitted precisely. Even when studying one's own consciousness, it is generally attempted to describe it in third-person perspective based on brain functioning, especially our senses. From a first-person perspective, we experience our perceptions, feelings, and thoughts in terms of qualia. We have seen how phenomenology performs part of this study.

We access the world by observing events and processes in the third person that are accessible to any observer and that we have placed ``outside" from our earliest years of life: from a solar eclipse to the fall of a fruit. Physics provides a mathematical description of the natural world in third person. It determines the regularities that any observer will notice. The regularism we have previously introduced establishes that regardless of how we account for the totality of observed phenomena including consciousness, the regularities ---the fundamental physical laws--- are always fulfilled without exception. Reducing consciousness to physics is unnecessary and probably impossible, but having detailed information about the behavior of the natural world in third person provides a starting point to understand the physical basis of consciousness.

As physics advances, incorporating new phenomena, the role of events becomes increasingly significant. Galileo determined that the distance traveled by an object falling down an inclined plane grows with the square of the time elapsed by observing coincidence events between the body and a ruler placed alongside the fall trajectory, correlating the elapsed times from launch and the distances covered. However, Newtonian mechanics does not consider events as fundamental, but rather focuses on material particles as the fundamental entities. The notion of an event occurring at a certain place at a certain instant becomes more central as physics develops, incorporating new phenomena. The first major change occurs with the advent of electromagnetism. It demands overcoming mechanical biases in trying to reduce phenomena to particle movements and accepting the physical existence of fields. 

This new vision relies, according to Laue \cite{laue} ``in a totally renewed understanding of the propagation of electromagnetic effects in vacuum; they are not supported by any medium nor do they give rise to an immediate action at a distance. The electromagnetic field in vacuum is something that has its own existence and an independent reality of any substance. Indeed, one must accustom oneself to this idea, but maybe this can be simplified if one remembers that the physical properties of this field, which are best given by Maxwell’s equations, are more perfectly and exactly known than the properties of any substance.” The new electromagnetic theory recognizes something as real if it can act on other entities and be modified by them. This change in the concept of a material substance did not go unnoticed among physicists and philosophers of science. For instance, more than a century ago Helmholtz said: ``With respect to the properties of the objects of the exterior world, it is easy to see that all properties we can assign to them mean only the effects that they produce on our senses or on other natural objects... In all places we occupy ourselves with the mutual relationships between bodies... From this we conclude that in fact, the properties of objects of nature are not, in spite of their name, nothing ``proper” of the objects themselves, in and of themselves, but they are always a relation to a second object (including the organs of our senses). The type of effect will depend, naturally, on the peculiarities of the body that produces it as well as those of the body on which the effect is produced.” \cite{helmholtz}. But to renounce to absolute properties of bodies ``in and of themselves” does not imply abandoning the objectiveness of knowledge that is not based on the absolute nature of things but in the events that they produce when they interact with a second object. Thus, the fields are defined by their effects, for instance, on test charges, magnets, or photographic plates, and will produce events by accelerating charges or creating an image on a photographic plate. Here, we encounter for the first time the concept of a reality consisting of events. As we put it in our book \cite{hospitable}: The relevance of events was reinforced by relativity and the quantum theory. In relativistic physics events are considered points in space-time: they occur at a certain place at certain instant. The events of the relativistic universe combine into partially ordered sets. An event B is in the future of another event A if it belongs to its future region defined by the light cone with vertex in A. In that case, A can influence B. If B is outside the future light cone of A, then it is causally disconnected from A, since the maximum speed of propagation of a signal is the speed of light. In general relativity light rays follow geodesics and the notion of past and future is dynamical, depends on the distribution of energy, and so does the notion of causality. 

Events take a central role with the advent of quantum mechanics at the beginning of the twentieth century. The new theory extends the description of nature to a new sphere of reality: the microscopic. It includes atomic and molecular phenomena, among others. Biological life at the molecular level depends strongly on the quantum properties of certain large molecules called proteins. Since large objects are composed of atoms or molecules, quantum mechanics is the most fundamental theory, its effects transcend the microscopic world, and it is necessary, for instance, for understanding the physical properties of solid and liquid materials. Our current understanding is that classical physics is only an approximation. The world, as described in the third person, is fundamentally quantum in nature. To understand the properties of the objects that surround us, one needs to resort to quantum theory. For instance, if we wish to determine the electric or thermal conductivity or the color that we will observe in a copper cable, we must resort to quantum theory \cite{persons}. 

The unification process of physics has not been completed. We have not yet been able to unify quantum mechanics with general relativity. Although we do not yet have a complete theory, we know that it will be quantum. In what follows, especially when we discuss the combination problem, we will show that quantum mechanics provides the ontology that best fits our phenomenal experiences. The argument that is usually used to set aside an analysis based on quantum ontology is that quantum mechanics cannot play any role, beyond biochemistry, in a medium like our brain. We will return to this point to show that this hypothesis is not justified by our current knowledge.

In quantum mechanics, the formalism refers to primitive concepts like a system, state, events, and the properties that characterize them. The use of these concepts suggests that the theory should admit an ontology that includes these concepts. A quantum system is described by a Hilbert space that represents the set of its possible states, which are vectors in it, and the events that may occur in the system which are represented by operators called projectors that act on that space and have information about all the their properties. Physical objects or simply objects and events can be considered the building blocks of reality. Physical objects are represented in the quantum formalism by systems in certain states. Given an object, we know all its possible behaviors when it interacts with other objects. Each behavior is known as an event. Events are the actual entities: the ones accessible to our senses about which we know in third person. The formalism of quantum mechanics associates a mathematical object, called a projector, with each event and its properties. This is important because the fundamental elements have a precise mathematical description. The element hydrogen is a quantum system. A particular atom of hydrogen is a system in a particular state. It is an example of what we call a physical object \cite{eventontology}. It is characterized by its disposition to produce events on other systems: for instance, the emission of a photon that produces a click in a photodetector with a certain probability. 

Note that objects are only observable through their manifestations in events resulting from their interaction with other objects. This does not mean that objects are mere hypothetical entities. We can operate on an object, for example, a hydrogen atom, and place it in a state that interests us, such as placing it in an excited state whose energy we know. But these operations are always performed based on information supplied by events: the unique entities accessible in third person to our consciousness. Quantum mechanics allows us, knowing the initial dispositional state of an atom, to calculate the possible energies with which a photon can be emitted and with what probability each transition will occur. Large objects such as a piece of iron are essentially quantum, although some of their behaviors, such as their macroscopic movement, can be described by Newtonian mechanics. Their visible appearance results from the events that occur on their surface when light falls on them.

The program of accounting for physical reality in terms of events goes back to Hume, who was the first to observe that a person's mind can be considered as a succession of events and a set of dispositions \cite{treatise}. In this regard, Russell says  \cite{matter} ``the enduring thing or object of common sense and the old physics must be interpreted as a world-line, a causally related sequence of events, and ... it is events and not substances that we perceive.” To put it differently, for Russell, an object is nothing more than a set of events that are causally connected. Although we consider this view to be a step in the right direction, we think it is insufficient to describe physical reality in terms of events, particularly in light of quantum mechanics. In fact, the causal connection between events is only understandable in terms of systems that are in certain states, what we have called objects, that evolve obeying the Schr"odinger equation.  As we put it in \cite{eventontology}: The basic idea of a measurement is the occurrence of a macroscopic phenomenon, that is, something capable of reaching perception. Thus, as noted by Omnes \cite{Omnes}, the measurement of a property of a microscopic object implies making it generate a phenomenon, in other terms, produce an event. The process of detection of photons by dissociation of silver bromide in a photographic plate leading to a cascade effect that produces the accumulation of millions of atoms of silver is an example of the production of an event. The appearance of a dot on the photographic plate is an example of a macroscopic event that constitutes the world accessible to our senses. 

Quantum field theory allows us to demonstrate that the electromagnetic field and others also admit an ontology of objects and events, and the existing versions of quantum gravity that unify general relativity with quantum mechanics show that the same ontology is applicable to spacetime.

We are interested in analyzing how far consciousness can be understood from a third-person perspective, such as that provided by science. In particular, is it possible that a physical approach to brain functioning might be at least compatible with the existence of a phenomenal world, such as the one perceived and phenomenologically analyzable? Russell thought that this compatibility between physics and what is perceived is essential. ``Hence if modern physics invalidates perception as a source of knowledge about the external world, and yet depends upon perception, that is a valid argument against modern physics. I do not say that physics in fact has this defect, but I do say that a considerable labour of interpretation is necessary to show that it can be absolved in this respect. And it is because of the abstractness of physics, as developed by mathematicians, that this labor is required.” A physical description of the brain must at least explain how the knowledge of the brain in the third person is compatible with our conscious activity.

\subsection{Panprotopsychism and the quantum ontology}

Husserl emphasized that psychology strives to approach with the techniques of other natural sciences that deal with objective phenomena a sphere of reality that can only be analyzed in the first person, thus distorting the essential character and meaning of human subjectivity and intersubjectivity. quote{crisis}. As noted by Chalmers, the definition of consciousness given in the International Dictionary of Psychology cannot be more ambiguous: ``Consciousness: the having of perceptions, thoughts and feelings; awareness. The term is impossible to define except in terms that are unintelligible without a grasp of what consciousness means... Nothing worth reading has been written about it.”. 
He is more explicit in what he pretends from a theory of consciousness\cite{chalmers}; he says ``Ultimately, one would like a theory of consciousness to do at least the following: it should give the conditions under which physical processes give rise to consciousness, and for those processes that give rise to consciousness, it should specify just what sort of experience is associated. And we would like the theory to explain how it arises, so that the emergence of consciousness seems intelligible rather than magical. In the end, we would like the theory to enable us to see consciousness as an integral part of the natural world. Currently it may be hard to see what such a theory would be like, but without such a theory we could not be said to fully understand consciousness.” 

In our understanding, his last requirement, having a theory that would enable us to see consciousness as an integral part of the natural world, implies a premise that makes the proposed objective unachievable. Natural sciences, as we have repeatedly emphasized throughout the work, provide a description of the world in the third person. They explain the world based on the information that reaches our senses about its behavior and its reaction to the experimental operations we perform on it. Consciousness is the manifestation of our direct access in the first person. At most, we could only hope to understand which states and events in the brain correspond to certain perceptions and feelings.

As we emphasized above, the world has a fundamentally quantum nature; concrete entities accessible to our senses are constituted by events localized in space-time that obey the laws of quantum mechanics. Quantum ontology includes, as we have discussed, objects, that is, systems in defined states, and events. We will initially limit ourselves to discussing the role of quantum events. That is, observable entities that occupy a small region of space-time, as discussed, for instance, by Omnes \cite{Omnes}. Whitehead \cite{whitehead} considered them as: `` the ultimate unit of natural occurrence.” Events come with associated properties whose mathematical representation is given by projectors, that is, a particular kind of operator in the Hilbert space of the system under observation. The event ontology we have presented has the attractive feature of eliminating the divide between the mental and the material world. As Russell \cite{mind} pointed out ``if we can construct a theory for the physical world which makes its events continuous to perception, we have improved the metaphysical status of physics” According to his view we need “an interpretation of physics which gives a due place to perceptions.” 

As we put it in \cite{hospitable}: An ontology of events could provide this interpretation: events in the external world are subject to a physical description, while at least some events in our brain could be directly accessible as perceptions. The main difference between both forms of events is the way we access them: first-person access for the mental, and third-person access for the physical.  This is a form of panprotopsychism, which is the doctrine that fundamental physical entities, such as events and states, have intrinsic features that, although not mental, when appropriate arranged, give rise to consciousness in complex creatures like us \cite{wishom}. 

In a quantum ontology, conscious phenomena such as sensations could be associated with events in our brain to which we have first-person access. In other words, whereas we only have indirect third-person access to the events of the systems we study physically, that is, we know of them ultimately by their effects on other objects and on us, we have direct access to the events in our brain, as long as we are aware of them. In that sense, we can say that we only have an idea of how things are in themselves through our conscience. One could say that the proposed panprotopsychism here is a kind of protophenomenal regularism. It is protophenomenal because the fundamental physical entities would have phenomenal intrinsic aspects of a primordial kind, which do not necessarily would entail a subject with a first-person perspective, and it is a regularist view because the third-person aspect of the fundamental entities only requires their causal connection with other entities obeying physical laws. Bertrand Russell adopted in 1927 \cite{matter} a position close to this form of panprotopsychism. He says: `Matter in a given place are all the events that are there...” and he goes on to say in the same book that such vision of matter implies that we ``do not have to deal anymore with what used to be mysterious about the causal theory of perception: how a series of waves of light or sound produce an event apparently totally different from them in its character.” More explicitly, he adds ``I think... that an ultimate scientific account of what goes on in the world, if it were ascertainable, would resemble psychology rather than physics... such an account would not be content to speak, even formally, as though matter, which is a logical fiction, were the ultimate reality. I think that, if our scientific knowledge were adequate to the task, which it neither is, nor is likely to become, it would... state the causal laws of the world in terms of... particulars, not in terms of matter. Causal laws so stated would, I believe, be applicable to psychology and physics equally; the science in which they were stated would succeed in achieving what metaphysics has vainly attempted, namely a unified account of what really happens, wholly true even if not the whole of truth, and free from all convenient fictions or unwarrantable assumptions of metaphysical entities.”\cite{mind}. It is this description in terms of particulars that could allow simultaneously having a third-person perspective obeying physical laws and an intrinsic phenomenal aspect. Although the position presented here shares strong similarities with the one proposed by Russell, it differs in several aspects that we will discuss next. In particular, between the following two alternatives: i) the hidden nature of physical reality is phenomenal in itself, the non-conscious building blocks of consciousness, or ii) building blocks that are neither physical nor mental, but which give rise to both \cite{wishom}. Russell seems to lean towards the second, known as neutral monism. However, as we have observed, natural sciences only access the world in third person and we only access our conscious mental contents in first person. Therefore, the first option that affirms the centrality of the phenomenal world seems to us the most suitable.

The quantum ontology differs from the notion proposed by Russell in that it includes, in addition to events, systems in certain states. If the quantum events have a phenomenal intrinsic aspect, does something similar occur with the systems in certain states that we have called objects?  What we describe as passions, emotions and the resulting voluntary behavior could be related with what we have described up to now as dispositional quantum states. Precisely this volitive aspect is what could be associated with states that in quantum mechanics have been related with dispositions or tendencies. Notice the similarities that exist between the characteristics of mental phenomena, such as emotions, intentions or feelings, dispositional and private, and those of the states. The latter are dispositional and inaccessible by isolated measurements, that is, they are private, and they may have an internal aspect like the one we assume here. As we have emphasized in \cite{hospitable}: states characterize the disposition of the system to act producing certain effects on other systems. To characterize the disposition of a state to act on any other system is to give its most complete description given by a vector or a density operator in a Hilbert space. States are private in the sense that one cannot determine precisely in which state a system is by measuring it. One needs an ensemble of identical states and measurements on each member of the ensemble in order to have a complete determination of the states. As it does not make sense to have an ensemble of identical mental states, they are inaccessible to external observers. It is important to note that states would simultaneously have psychological and phenomenal aspects. Indeed, it is the state that dictate the tendencies or dispositions that physics studies, which lead to the production of events. Quantum mental states fulfill a double function: Insofar as dispositions determine probabilities for certain actions, which are studied by psychology, and insofar as they possess an internal aspect expressed in desires or emotions, they are phenomenal.

Since quantum ontology has a probabilistic character, strict compliance with the regularities established by its axioms allows us to distinguish the regularism adopted here from the usual physicalism. Classical physics, from Newtonian mechanics to general relativity, is deterministic. For example, in Newtonian mechanics, once the positions and velocities of the particles that make up a system at a given instant and the forces with which they interact are completely determined, there is no possibility that a free action can alter its development without violating the laws of Newton’s mechanics. Mechanist determinism pushed in that way to its last consequences left the world without any possible opening to novelty. In classical physics, there is ``causal closure". In other words, that point of view stems from the conviction that every physical effect has a physical cause, and therefore everything that has physical effects is physical.
If physics were fundamentally classical, the regularist vision proposed above that only requires that the regularities that physics identifies in its description of processes and phenomena must be fulfilled without exception would be entirely equivalent to physicalism. If there is causal closure, then there are aspects of conscious activity such as aesthetics and ethics discussed in section 1 that must be formulated in terms compatible with determinism. Being quantum mechanics probabilistic, it is not closed: The occurrence of an event is not determined by previous causes but is limited only by its probability of occurrence. We will analyze in more detail in the next section what the freedom allowed by quantum mechanics may imply.

But the most important reason to consider a quantum treatment of consciousness is that it would allow us to solve the combination problem. This is the main objection to panpsychism or panprotopsychism. When thinking of matter composed of microscopic entities, the question arises: ``How any combination of more basic minds or mind-like qualities could somehow add up to conscious mental lives like ours” \cite{wishom}. William James in The Principles of Psychology \cite{james} described it this way: ``Where the elemental units [which compose complex minds] are supposed to be feelings, the case is in no way altered. Take a hundred of them, shuffle them and pack them as close together as you can (whatever that may mean); still each remains the same feeling it always was, shut in its own skin, windowless, ignorant of what the other feelings are and mean. There would be a hundred-and-first feeling there, if, when a group or series of such feelings were set up, a consciousness belonging to the group as such should emerge... but they would have no substantial identity with it, nor it with them, and one could never deduce the one from the others, or (in any intelligible sense) say that they evolved it.... Private minds do not agglomerate into a higher compound mind.''  The issue turns out to be how elementary micro-experiences combine to generate our complex mental experiences. The solution provided by quantum mechanics is satisfactory, and we will describe it in detail in the next section. In summary, when microscopic quantum objects interact, the states of the constituent objects, such as atoms, get entangled and their parts do not allow one to determine the state of the overall system. This is a form of quantum entanglement that has been analyzed and understood by many. The events and states of such entangled systems are produced by the complete system and not its parts. We call ``protophenomenal regularism" the form of panprotopsychism sketched so far. The considerations presented above, about which we will return in detail in the next section, point against the widely held opinion that brain function can be understood based on the exchange of signals between neurons, towards the existence of a quantum system coupled with the neural one.

However, there are physical objections that every model of consciousness must face. Some years ago, Tegmark \cite{tegmark} presented a quantitative analysis of the effects of environmental decoherence in the brain, suggesting that such processes occur very rapidly. The decoherence timescales are typically much shorter than the relevant timescales for regular neuron firing ---and will destroy entanglement---. In the next section, we will mention some models that allow us to deal with that objection. It should be noticed that in the last decades several manifestations of quantum mechanics, such as tunneling, superposition, coherence, and entanglement, have been found in different biological phenomena \cite{mcfadden}, \cite{ikeyawoodward}. Therefore, the decoherence effects observed by Tegmark are not sufficient to eliminate all quantum effects in biological systems beyond the molecular level.

\section{Quantum emergence of states and properties and the combination problem}

The idea of invoking quantum mechanics to explain consciousness in physical terms arouses skepticism among most scientists. It is generally considered that the only reason for this is to link two seemingly complex or mysterious problems. It contributes to skepticism the presumption that quantum mechanics cannot play any role in a humid and warm medium like the brain. In the section above, we argued that there are strong reasons to formulate the problem of consciousness using two of the fundamental properties of quantum mechanics: its ontology and quantum entanglement. However, the assumption that quantum mechanics plays no other role in biological systems than in the determination of molecular structures has been shown to be invalid. Quantum phenomena such as the tunnel effect or entanglement seem to fulfill an important function in photosynthesis, olfaction, or bird orientation. Regarding brain function, Adams and Petruccione \cite{ap} recently published a review that analyzes several possible quantum effects involved in the brain: ``the firing of nerves; the actions of anesthesia, neurotransmitters, and other drugs; the sensory interpretation and organized signaling that are central to the vast neural network that we identify as our self". Particularly interesting is Fisher's model of quantum cognition \cite{fisher}.

We will refer in what follows to the non-locality and entanglement of quantum states, effects that allow answering various questions linked to the combination problem. They arise from a mistaken conception of the functioning of matter within a panpsychist framework. A conception that can be summarized in the belief, based on a prejudice coming from classical physics, that a system of many particles in a certain state is nothing more than a set of individual systems in certain states and is false for quantum systems.  James summarizes this problem as follows ``Private minds do not agglomerate into a higher compound mind" \cite{james}. Within a classical framework, ``we do not have an intelligible picture of how the qualitative aspects of microexperiences could combine so as to yield the rich and varied macroexperiences of the sort we and other sentient organisms enjoy. Consequently, the emergence of our conscious mental lives from more basic natural elements (arguably) looks to be just as miraculous and inexplicable for panpsychism as it is for competing views" \cite{wishom}.
We will see that nonlocality and quantum entanglement contribute to understanding how protophenomenal properties combine to give rise to conscious mental contents.
 
The unity of consciousness that draws attention in different versions of the combination problem seems difficult to explain in terms of classical physics. In fact, classical properties are always Humean, as a mosaic. If one adopts a physicalist viewpoint based on classical physics, as is frequently done, the whole is the sum of its parts. The lower level contains all the information, and the higher level, consciousness, for example, must emerge from these parts. Any classical magnitude ---for example, total energy--- of a mechanical system of particles that interact with known forces is a function of the properties of the particles that make up the system: basically, their positions and velocities. In three dimensions, three variables are needed to define position and three for velocity. If there are N particles, there are 6N variables available to define any macroscopic quantity. Therefore, the palette of microqualities would be quite reduced. We will see below that conversely, in quantum mechanics, the number of independent observables grows exponentially with the number of particles due to the entanglement or non-locality of quantum systems.

\subsection{Non-locality, non-separability, and
quantum emergence}

The role of entanglement and non-separability of quantum states was only recognized in the 1930s. During this period, two fundamental works were published. Einstein, Podolsky, and Rosen argued, using an entangled system, that quantum mechanics was apparently incomplete. Schrödinger then published a work stating that entanglement was crucial: ``[it] is not one, but rather the characteristic trait of quantum mechanics, the one that enforces its entire departure from classical lines of thought."
In Tim Maudlin's work ``Why be Humean?" \cite{maudlin} he clearly demonstrates how quantum mechanics alters physical and philosophical conceptions about the local nature of reality resulting from Newtonian mechanics and empiricism. He begins by recalling the Humean version of supervenience. ``We have geometry: a system of external relations of spatio-temporal distances between points... And at those points we have local qualities: perfectly natural intrinsic properties which need nothing bigger than a point at which to be instantiated. For short, we have an arrangement of qualities. And that is all. There is no difference without difference in the arrangement of qualities. All else supervenes on that" \cite{lewis}.

This is a concept that has been completely changed by quantum mechanics. As we put it in \cite{strong}: Quantum systems may be in certain quantum states, called entangled, which have well-defined properties that follow neither from the properties of parts nor from relations among them. Most of the properties of a system do not have well-defined values until measured; for instance, the position of an electron in the double-slit experiment is not well defined until a dot is produced in the photographic plate and it is detected. Although, in general, properties cannot be attributed to states, quantum systems in a pure state have some well-defined properties. To exemplify this, let us consider a spinning particle like the electron. While a classical particle rotating along an internal axis may have a continuum of values for the projection of its angular momentum on a given direction of space, in quantum mechanics its projections along an arbitrary axis are quantized. For example, a spinning particle such as an electron can only take two possible values of its projection along an axis $z$: up or down. Given a spinning particle such as the electron, one can measure its spin projection along $z$ using a Stern--Gerlach device. It mainly consists of a magnet with S-N poles oriented along the axis in question, $z$, in this case, and a photographic plate. When repeated measurements are performed on particles in a generic state, dots appear in the upper region with a certain probability and in the lower region with complementary probability. When the electron is in a state that leads with certainty to a dot in the upper region, it can be said to be in the state $\vert z up\rangle$. In this case, the property of pointing along $z$ in the $up$ direction may be assigned to the state. This is the only property that can be assigned to this state. Measurement of any other projection of the spin along a different direction will not lead to a unique value, i.e. always up or always down. It is only when one knows with certainty what the behavior of the system in a certain state will be that one may assign a property to the state.

As indicated in quantum mechanics, there exists a stronger form of emergence that differs from the Humean form of supervenience, where the properties of the whole do not arise from the properties of its parts or their relationships. This happens in entangled systems, that is, in most quantum systems. 
Let us see an example of entangled states in a simple system: two particles with spin. Let us assume that the two particles are prepared in two laboratories with spins that point in the same $z$ direction. In laboratory $1$ the particle is in the state $\vert 1, z, up\rangle$ and in $2$ is in the state $\vert 2, z, down\rangle$. One can think of the two-particle system as a single physical system in a state $\vert \psi\rangle = |1, z, up\rangle \vert 2, z, down\rangle$. These types of states are known as tensor products and represent particles that basically behave as independent. The effects and predictions about measurements made on one of them would be the same as those that would result if the other did not exist. For example, the probability of obtaining a measurement of the $x$ component of the spin of particle $2$ is $1/2$. But not all two-particle systems behave in this trivial manner. The superposition principle says that any superposition of states of the two particles is a possible state of the system. For example, a possible state of the system of the two particles, and our first example of an entangled state, is called the triplet state,
\begin{equation}
|\psi_T \rangle = 1/ \sqrt{2}|1, z, up\rangle|2, z, down\rangle + 1/\sqrt{2}|1, z, down\rangle|2, z, up\rangle. 
\end{equation}
that represents a system of two particles with a defined total spin of 
$\sqrt{2}\hbar$  and a total $z$ component equal to $0$. Although the total system has defined properties, it is not possible to assign a definite spin orientation ---with probability one--- to one of the particles that make it up. Regardless of the direction we measure the spin, half the time it will be up and half the time it will be down.  The total system has properties --—the $z$ component of the spin—-- that do not arise from the properties of its parts. In fact, the particles that compose it do not have any well-defined property ---component of the spin---. Therefore, no pure state of particle $1$ will give the answers that it gives when entangled. The same happens with particle $2$.

Notice that there exists a second entangled state obtained by considering the difference of the same tensor products, which we shall call the singlet state, given by: 
\begin{equation}
|\psi_S \rangle = 1/ \sqrt{2}|1, z, up\rangle|2, z, down\rangle - 1/\sqrt{2}|1, z, down\rangle|2, z, up\rangle. 
\end{equation}
It represents a system of two particles with total spin $0$ and vanishing $z$ component. The analysis we carried out regarding the answers that $\vert \psi_T \rangle$ would give to measurements performed on one of the particles also holds for $\vert \psi_S \rangle$. The prediction for the probability of finding the spin of one of the particles pointing $up$ in any direction when the system is in such a state is always $1/2$. The quantum formalism allows us to extend the notion of state and describe one of the particles of the entangled pair, even if it is not in a pure state because it does not have any well-defined property. We saw that no state with spin in a given direction can describe one of these particles. To describe the local results that an observer would get taking measurements on one of the particles of the pair, one must generalize the notion of state and introduce the so-called density operators. In this case, the density operator is a mixture state. The density operator of particle one would be called $\rho_1$ and would correspond to $\vert 1, z, up\rangle$ with probability $1/2$ and $\vert 1, z, down\rangle$ with probability $1/2$.  One can prove that any quantum state, including pure states, may be written in terms of density operators, which are the most general description of a state.

To distinguish between the triplet and singlet states, one must measure the entire system or study the correlations between the measurements made at $1$ and those made at $2$. Various authors have noted that the existence of entangled states shows that the local notion of supervenience is inapplicable to quantum systems. As shown in the previous example in quantum mechanics, there exists what Healey \cite{healey} calls ``Physical Property Holism". It establishes that there are physical objects ``not all of whose qualitative intrinsic physical properties and relations supervene on qualitative intrinsic physical properties and relations in the supervenience basis of their basic physical parts.” The emergence of new properties of the whole in a quantum world, where events and properties play a fundamental role, is a crucial manifestation of the ontological novelty and non-separability most characteristic of quantum emergent phenomena. These behaviors of entangled systems allow wholes to have properties that cannot be explained in terms of the properties of their parts. In the model of two entangled spinning particles, the total spin of the two-particle system is an example of a property that does not supervene from its parts. In fact, the parts do not have any defined component of the spin. The novelty of these properties can play an essential role in the emergence of consciousness in composite systems with a large number of components.

Paul Teller \cite{teller} has given a different
characterization of the behavior of entangled states. He defines as local supervenience the following feature of any classical system: ``the world is composed of diverse physical individuals that possess non-relational (i.e., monadic) properties; and... all relations existing among such entities supervene upon the non-relational properties of the relata.” He defines as ``relational holism” the stand point that admits that there exist relations that do not supervene from the non-relational properties of their relata. Entangled states in quantum mechanics are of this kind; that is, they exhibit relational holism. In the example of entangled spins, in fact, there are no non-relational properties associated with the spin components of their relata.

Karakostas \cite{karakostas} compares the resulting quantum picture with the Humean ontology as follows: ``From the perspective of Lewis’---Humean--- metaphysic, if one is able to determine the intrinsic qualities of particular events or atomic objects in space and time, then one can describe the world completely. Quantum mechanics, however, is not in conformity with Lewis’ atomistic metaphysical picture that depicts a world of self-contained, unconnected particulars that exist independently of each other... the consideration of physical reality cannot be comprehended as the sum of its parts in conjunction with the spatiotemporal relations among the parts, since the quantum whole provides the framework for the existence of the parts... their entangled relation does not supervene upon any intrinsic or relational properties of the parts taken separately. This is indeed the feature that makes the quantum theory go beyond any mechanistic or atomistic thinking.” The quantum world is essentially non-separable. As we shall see below, this is true not only for properties but also for states.  The ``holistic" behavior of many quantum systems that are non-separable is not something exceptional. It is the most common occurrence in composite systems whose components interact or have interacted in the past. For example, the electrons in a multielectron atom are entangled. It is a generic property: interactions typically yield entangled states for multi-component systems. 

Unlike in classical physics, where the number of possible states of motion grows linearly, the space of possible quantum states of a composite system of particles grows exponentially with the number of particles. In systems of $n$ particles, because the dimensionality of the space of states of the system increases exponentially with $n$. For example, for $n$ spins, the number of independent states grows as $2^n$. Non-separability and holism become more important the more particles the systems involve. This exponential growth in the possible behaviors of the quantum systems and the possible entangled states, is at the basis of the appearance of novel properties and top-down behaviors of emergent systems. Their philosophical implications can be recognized when the appropriate ontology is put into action. For instance, most of the events, described by projectors in Hilbert space, correspond to entangled states and differ from the simple sets of events of the non-entangled parts. The assumption that it is not necessary to analyze the mind-body problem from a quantum ontology is not well founded. First, because there are quantum models of cognition compatible with brain physical conditions, and secondly because it cannot be ignored that quantum mechanics provides an ontology much more adjusted to the phenomenic evidence.

Another typical characteristic of quantum systems is the emergence of novel causal properties. A system will present top-down causation if the parts have some behaviors that are dictated by the state of the whole and that cannot be predicted from the knowledge of the state of the parts: there is state non-separability. As we discussed in \cite{strong}: The previous example of an entangled state shows that in quantum mechanics there is state non-separability. The states of the parts are just a statistical mixture of up-components and down-components, while the complete entangled state has more information. It is this state non-separability that leads to top-down causation. Let us study this behavior in some detail. 
Let us start by recalling that for each entangled pure state, such as the ones shown above, there is a corresponding property given by a projector. For example, $\vert\psi_S\rangle\langle\psi_S\vert$ is the property of a system of two particles with vanishing spin.
It coincides with the density operator that describes the state of the entangled system. In fact, a pure system may be described by a vector $|\psi\rangle$ or by its density operator given by the projector $\rho=|\psi\rangle\langle\psi|$. 
Let us assume that George prepares the entangled state $\vert \psi_S\rangle$ we discussed before and sends the particle $1$ to Alice and the particle $2$ to Bob. We will assume that both have devices that allow them to measure spin projections along $x$ or $y$. Without any communication, Alice and Bob choose independently and in a random fashion to measure one of these projections every time they receive a particle. If the system is in this state, it does not matter which component they measure, they will have probability $1/2$ of measuring ``up” and $1/2$ ``down”. However, if they compare notes, they will realize that the sequences will be correlated: whenever they happen to measure the spin in the same direction their results will be opposite, up for Alice and down for Bob or vice versa. They could never have figured this out by looking at the individual systems in isolation. In fact, the state of the system accessible to Alice is nothing but the density operator defined above for the particle $1$, $\rho_1$.  The complete system has a certain non-locality such that when one electron chose to answer ``up”, the other necessarily needs to choose ``down”. Such a correlation does not involve time, it is instantaneous and is not predicted by the states of particle $1$ or particle $2$, but results from the state of the complete entangled system.  Contrary to what happens in classical physics, the state of the whole system is not determined by the states of its parts. There is top-down causation. Mathematically $\rho_S\ne\rho_1\otimes\rho_2$, and similarly for the state $\rho_T$ \cite{strong}.

Even though local supervenience is not valid in quantum mechanics, there is an extended notion that should be called {\em quantum emergence of properties} that could be defined as follows: The properties of a quantum system $U$ composed of $n$ subsystems with Hilbert spaces $H_1,\ldots,H_n$ are given by the projectors $P_U$ defined in the tensor product Hilbert space $H_1\otimes H_2 \ldots \otimes 
H_n$. When the subsystems have defined properties, these projectors take the form $P_1\otimes P_2 \otimes \ldots P_n$ and the properties are represented by projectors in $H_1\otimes H_2\otimes \ldots H_n$, but most of the properties ---projectors--- in this space are not tensor products of the properties of $P_1, P_2,\ldots P_n$. They are emergent properties of entangled states, such as the ones discussed above, for a system of two particles.
There is an analogous notion of {\em quantum emergence of states}
that we shall not discuss in detail, but is closely related to entanglement and can be easily derived from the above definition of properties by substituting projectors by density operators.

\subsection{Quantum emergence and the combination problem for panprotopsychism }

David Chalmers in his recent paper The Combination Problem for Panpsychism \cite{chalmers2}  attempts at a systematic treatment of the combination problem. It is a rigorous analysis that simplifies the examination of its hypotheses. He analyzes several aspects or versions of the problem that raise objections against panpsychism. Most of these objections result, as we shall see, from an implicit use of notions, like the traditional idea of local supervenience, valid in classical physics but superseded by quantum mechanics. 

He starts with some terminological issues for panpsychism, which we reproduce here: ``Microphysical properties and entities are the fundamental physical properties and entities characterized by a completed physics. Phenomenal properties are properties characterizing what it is like to be a conscious subject. Microphenomenal properties are the phenomenal properties of microphysical entities. Macrophenomenal properties are the phenomenal properties of other entities, such as humans. Microphenomenal and macrophenomenal truths are the truths about the instantiation of these properties. ” For the case of panprotopsychism, he adds the following comment: ``Panprotopsychism is the thesis that fundamental physical entities have protophenomenal properties. Protophenomenal properties are special properties that are not themselves phenomenal (there is nothing it is like to have them) but that can collectively constitute phenomenal properties. To rule out standard forms of materialism from counting as panprotopsychism, these special properties must be (i) distinct from the structural/dispositional properties of microphysics and (ii) their constitutive relation to phenomenal properties must reflect an a priori entailment from protophenomenal to phenomenal truths.” He finally defines ``Constitutive panpsychism is the thesis that macrophenomenal truths are (wholly or partially) grounded in microphenomenal truths. Non-constitutive panpsychism is the thesis that macrophenomenal truths are not grounded in microphenomenal truths. The most important form of non-constitutive panpsychism is emergent panpsychism, on which macrophenomenal properties are strongly emergent from microphenomenal or microphysical properties, perhaps in virtue of fundamental laws connecting microphenomenal to macrophenomenal.” 

When judging the appropriateness of these concepts, the judgment radically changes depending on whether one assumes that the physical foundations of consciousness are classical or quantum. Chalmers claims that if one adopts the constitutive panpsychism: ``microphenomenal properties are causally efficacious in virtue of their playing fundamental microphysical roles, and macrophenomenal properties are causally efficacious in virtue of being grounded in microphenomenal properties. By contrast, non-constitutive... panpsychism has many of the same problems with mental causation as dualism.”
This statement implicitly carries the notion originating from classical physics that the state of a system is the union of the states of its parts and, therefore, the behavior of the whole is completely determined by the behavior of its parts. We have seen that this is not the case in quantum systems in entangled states, where the states of the parts are not causally effective in explaining many of the behaviors of the whole, leading to what we have called top-down causation. If there were quantum phenomena involving entanglement, they would lead to macrophenomenal properties strongly emerging from microphenomenal or microphysical properties, and therefore if it were possible that conscious phenomena had a quantum origin, it should be adopted a form of non-constitutive emergent panpsychism.

To present the combination problem, Chalmers starts with a general statement which he then decomposes into various sub-problems. Succintly, the problem is posed in the following terms: ``How can protophenomenal properties combine to yield macrophenomenal properties?" The problem can be broken down into three sub-problems:

First, the subject combination problem, which in the case of panprotopsychism would involve understanding how elements with internal aspects lacking subjectivity can organize into subjects capable of first-person perspectives of the world and themselves. 

Second, the quality combination problem, according to Chalmers, would involve understanding ``how do microqualities combine to yield macroqualities? Here macroqualities are specific phenomenal qualities such as phenomenal redness (what it is like to see red), phenomenal greenness, and so on." Again, in the panprotopsychist version, it would seem that heterogeneous qualities are being compared. If the world behaved classically, the problem would be how qualities that accompany certain physical events, such as a neuron firing, yield the qualities that appear at the end of processes that start in our sensory organs and are organized by our brain into complex perceptions. However, the preceding quantum analysis of states and events allows the problem to be posed in much more promising terms. Indeed, as we have seen, the quantum emergence of states and properties does not allow us to think that the properties or states of the whole are composed of properties or states of the parts as in classical physics. Instead, in a quantum system, the parts lose their individuality and, as in the previous examples of spin systems, only the properties of the whole are defined. In other words, in a quantum system capable of maintaining its entanglement, macroqualities would appear without leaving a trace of the microqualities of the entities involved in its constitution. 

Third: the structural combination problem (sometimes called the ``grain problem"), which concerns how complex arrangements of vast numbers of discrete, spatially discontinuous insentient materials could yield a homogeneous experience. Like the experience of red or perceptions that are unified, multifaceted, and structured both egocentrically and in terms of objects represented as being in the environment.
The third subproblem refers to our ability to represent and interpret the reality, that we have identified with the intentionality of our perceptions and the construction of the life-world. Let us remember that both the organization of our perceptions and the construction of the life-world require an active process of the subject and its interactions with others and the world. The problem results from that incorrect prejudice in a quantum system of the supervenience of the state of the total system on the states of the microsystems. As we saw when analyzing top-down quantum causality, the union of individual entities does not represent the entire entity. The above mentioned homogeneous experience results from a process of elaboration involving large portions of the brain interacting with an entangled system of many particles. To enable a solution, like the suggested one with top-down causation, it is necessary to have a quantum system composed of many microscopic quantum components, such as molecules, in entangled states that extend across macroscopic regions of the brain and capable of coupling with the brain's neural system.

The most tangible test about whether that prejudice that assumes that the properties of the whole come from its parts, a hypothesis that is true in classical physics but false in quantum physics, is posed as another subproblem linked to the combination problem that Chalmers calls the palette problem: ``An especially pressing aspect of the quality combination problem is what we might call the palette problem. There is a vast array of macroqualities, including many different phenomenal colors, shapes, sounds, smells, and tastes. There is presumably only a limited palette of microqualities. Especially if Russellian panpsychism is true, we can expect only a handful of microqualities, corresponding to the handful of fundamental microphysical properties. How can this limited palette of microqualities combine to yield the vast array of macroqualities?" Here, clearly appear the prejudices resulting from classical physics that generate this problem. Indeed, again it is explicitly assumed in this case, corresponding to constitutive panpsychism, that the properties of the whole come from those of its parts, and therefore grows linearly with the number of parts. Quantum mechanics, by providing an exponential number of states and properties in which, in most of them, the properties of the parts are not defined any more, gives a satisfactory answer to the palette problem. The states and events in our brain can have the complexity of our emotions and perceptions.

Although many subproblems of the combination problem are easier to solve in a quantum context, many questions remain open. For example, in relation to the first sub-problem mentioned above: How elements with internal aspects that lack subjectivity can be organized into subjects or individuals who can process information from the environment that surrounds them, preserving memories and preferences, and act on the world by making decisions, as well as many other problems related to phenomenal and psychological aspects need to be investigated.

\subsection{Quantum models of the brain}

But what do we mean when we point out the need to include a quantum system capable of maintaining enough coherence and entanglement levels in the brain? Certainly not that the neural network, which plays an essential role in organizing and interpreting the information that reaches our senses and which neuroscience has determined to operate fundamentally based on the exchange of electromagnetic signals, should have any macroscopic quantum nature. The mere association of emergent conscious states, such as qualia with certain neural electromagnetic configurations, seems as unfounded as the attribution of collective will to the movements of a flock of birds. As in any classical system, only weak emergence ---in the sense of Bedau \cite{bedau2}--- would be allowed, and any property of the whole would supervene from the property of the parts and their relations. In physical terms, the phenomenal aspects of consciousness would supervene from the electromagnetic properties of neural circuits and their relations \cite{persons}. What we must seek is a system of many microparticles capable of sustaining a quantum level of activity in a warm and humid medium such as the brain and of coupling and interacting with the neural network.

Adams and Petruccione \cite{ap} have observed that it is perhaps misleading to talk about quantum processes limited to the brain. ``Nerve cells extend throughout the body. The quantum processes could take place within neurons and at synapses. They are therefore implicated in the biological functioning of the entire body. It might be more accurate to describe these effects as quantum enhanced neural processing.” One of the most attractive models in this context is the one proposed by Fisher for entangled Posner molecules \cite{fisherclever}. The hypothesis is that the nuclear spin of phosphorus might function as a neural qubit, allowing for quantum processing to play a role in cognition. 

Fisher \cite{fisherclever} proposed a model based on the following observation about the effects of lithium in tempering mania and bipolar disorder. Lithium ions Li+ are very simple atoms with a spherical electronic structure. This simple element has, however, remarkable effects on human ---and animal--- psyches. The two isotopes of lithium, Li-7 and Li-6, have opposite effects on the behavior of rats. For rats taking Li-7, the cognitive capabilities become slowed down while for Li-6, they were enhanced. But both isotopes have exactly the same chemical behavior. The origin of the different properties may only be attributed to their different nuclear properties. The most remarkable difference is their nuclear spin and therefore their capability of electromagnetically interacting with the environment. While Li-7 has a decoherence time of a few seconds, Li-6 has decoherence times of the order of several minutes. The apparent correlation between nuclear properties and cognition led Fisher to study possible mechanisms of quantum information processing in the brain based on quantum units of information associated with the nuclear spin. 

We will not enter into the details of Fisher’s proposal of entangled microsystems protected from decoherence effects. As we put it in \cite{physical}: summarizing, the model is based on nuclear spins capable of interacting with the neural system. The model describes the entanglement of phosphorus spins that are part of Posner clusters, also known as Posner molecules, that have large decoherence timmes. They are transported to presynaptic neurons (a neuron from the axon terminal of which an electrical impulse is transmitted across a synaptic cleft to the cell body or one or more dendrites of a postsynaptic neuron by the release of a chemical neurotransmitter).  Posner molecules eventually decompose,  producing postsynaptic neuron firing events. Multiple entangled Posner molecules, triggering non-local quantum correlations of neuron firing rates, would provide the key mechanism for neural quantum processing. He proposed a mechanism for transporting components of these multiple-entangled systems to different regions of the brain. Neuron firings can be triggered via quantum event measurements. The net result would be having nonlocal quantum correlations of neuron firings. The existence of entangled quantum systems in the brain capable of performing some form of quantum computations and interactions with neurons seems to be the main requirement for conscious systems that include effective forms of mental activity such as decision-making. Models like Fisher's are those that can provide the necessary elements for a conscious activity consistent with the observed phenomenal behavior. The search for this type of models should be approached urgently and systematically.

\section{Conclusions}

Consciousness, both in its psychological and phenomenal manifestations, is analyzable and has physical manifestations. Phenomenology allows us to analyze conscious awareness based on primary information provided by its phenomenal manifestations: our ways of perceiving, feeling, or desiring, and our access to a world toward which consciousness is always intentionally oriented. Physical and mathematical sciences provide a means to access observed regularities and transmit them intersubjectively through their articulation in precise mathematical laws. As such, they do not say anything about the intrinsic aspect of things or establish any other limits on them beyond those determined by the regularities that they fulfill in their causal relationships. In other words, in the processes in which they are deployed. Classic mechanistic determinism, being deterministic, limits everything that can occur to what is given in an instant and leaves no space for novelty or any intrinsic aspect that is not epiphenomenal. That is why the temptation arises to reduce the object to its mathematical description and fall into what Whitehead called``the attribution of misplaced concreteness," which was practically irresistible in the realm of classical mechanics. But this would imply that mental events are not part of the explanations of our actions \cite{honderich}. The situation changes with quantum probabilistic determinism if the ontology of objects and the events they give rise to by interacting with others is taken seriously. An ontology arises from its own axioms. As Bertrand Russell noted almost a century ago, an ontology of events, with an internal phenomenal aspect, now known as panprotopsychism, is more suitable for accounting for the phenomenal aspects of consciousness. The central observation of this paper is that many objections to panpsychism and panprotopsychism, which summarize the combination problem, arise from hypotheses based on classical physics about locality and local supervenience, which are inappropriate at the quantum level. Quantum mechanics and the resulting protophenomenal regularism here considered facilitates the treatment of the Combination Problem ---see Wishom \cite{wishom}--- and allows for a personal dimension that accommodates its conceptual ethical and aesthetic manifestations. At the same time, the analysis imposes conditions on the possible implementation of a mechanism of quantum cognition, making the problem of consciousness experimentally accessible.

\section{Acknowledgements}
This work was supported in part by grant  NSF-PHY-2206557, funds of the
Hearne Institute for Theoretical Physics, CCT-LSU, Fondo Clemente Estable
FCE 1 2019 1 155865.

\end{document}